\documentclass[%
 reprint,
showpacs,
 amsmath,amssymb,
 aps,
 pra,
longbibliography,
floatfix,
superscriptaddress
]{revtex4-1}

\usepackage[utf8]{inputenc}	
\usepackage[T1]{fontenc}	
\usepackage[]{graphicx}		
\usepackage{color}
\usepackage[english]{babel}
\usepackage{hyperref}
\usepackage{url}
\usepackage{todonotes}
\usepackage{lipsum}
\usepackage{bm}

\usepackage{microtype}
\usepackage{braket}
\usepackage{esvect}
\usepackage[export]{adjustbox}

\usepackage{mathtools}

\usepackage[position=top,caption=false]{subfig}

\let\originaleqref\eqref
\renewcommand{\eqref}{Eq.~\originaleqref}

\newcommand{\fref}[1]{Fig.~\ref{#1}}

\newcommand{\e}[1]{\mathrm{e}^{#1}}

\newcommand{\sx}[0]{\hat{\sigma}_\mathrm{x}}

\newcommand{\sz}[0]{\hat{\sigma}_\mathrm{z}}

\renewcommand{\H}{\hat{H}}

\newcommand{\I}{\mathcal{I}}
\newcommand{\Tm}{T_\mathrm{max}}
\newcommand{\M}{\hat{M}}
\newcommand{\psieq}{\psi_{\mathrm{eq}}}
\newcommand{\psiop}{\psi_{\mathrm{op}}}
\newcommand{\rhoss}[0]{\overline{\rho}}
\newcommand{\Slabel}[0]{S}
\newcommand{\Mlabel}[0]{M}

\makeatletter
\newcommand*\bigcdot{\mathpalette\cdot@{.5}}
\newcommand*\bigcdot@[2]{\mathbin{\vcenter{\hbox{\scalebox{#2}{$\m@th#1\bullet$}}}}}
\makeatother

\graphicspath{
{/home/alexander/Dropbox/PhD/thermometry/plots/}
}

\begin{document}
\title{A dynamical approach to ancilla assisted quantum thermometry} 

\author{Alexander Holm Kiilerich}
\email{kiilerich@phys.au.dk}
\affiliation{Department of Physics and Astronomy, Aarhus University, Ny Munkegade 120, 8000 Aarhus C, Denmark}

\author{Antonella De Pasquale}
\affiliation{NEST, Scuola Normale Superiore, and Istituto Nanoscienze CNR, 56127 Pisa, Italy}
\affiliation{Dipartimento di Fisica e Astroniomia, Universitá di Firenze, I-50019, Sesto Fiorentino (FI), Italy}
\affiliation{INFN Sezione di Firenze, via G.Sansone 1, I-50019 Sesto Fiorentino (FI), Italy}

\author{Vittorio Giovannetti}
\affiliation{NEST, Scuola Normale Superiore, and Istituto Nanoscienze CNR, 56127 Pisa, Italy}

\date{\today}
\bigskip

\begin{abstract}
A scheme for improving the sensitivity of quantum thermometry is proposed where the sensing quantum system used to recover the 
temperature of an external bath is dynamically coupled with an external ancilla (a meter) via a Hamiltonian term $\hat{H}_I$.
At variance with previous approaches, our scheme relies neither on the presence of initial entanglement between
the sensor and the meter, nor on the possibility of performing 
joint measurements on the two systems. The advantages we report 
arise from the fact that the presence of $\hat{H}_I$ interferes with the 
bath-sensor interaction, transforming the sensor into an effective transducer  
which extracts the intrinsically incoherent information on the bath temperature,  
and maps it into coherences in the meter where it can finally be recovered by local measurements.
\end{abstract}

\maketitle
\noindent

\section{Introduction}
The aim of thermometry is to estimate with high precision the temperature $T$ of a thermal bath, and a thermometer consists of a probe system which is put in contact with the bath of interest. By monitoring the state of the probe one seeks to recover the value of $T$. If the probe is small, this has the advantage of inducing a negligible disturbance to the thermal equilibrium of the reservoir.
The same principle applies in the quantum regime and substantial interest has recently been devoted to the design and properties of sensitive quantum thermometers \cite{de2016local,PhysRevLett.114.220405,PhysRevA.96.012316,campbell2018precision,PhysRevA.91.012331,PhysRevA.82.011611,PhysRevLett.119.090603,mukherjee2017high,hofer2017fundamental}.

By employing single or few-body quantum probes it has proven possible to obtain very precise temperature readings at millikelvin temperatures with spatial resolution at the nanometer scale. 
For instance, single quantum dots and NV-centers in nanodiamonds experience frequency shifts which depend on the temperature of their surroundings, thus allowing their implementation as sensitive fluorescent thermometers \cite{doi:10.1021/nl401216y,yang2011quantum,kucsko2013nanometre,toyli2013fluorescence,seilmeier2014optical,haupt2014single}. Other designs utilize mechanical oscillators or spin systems \cite{PhysRevA.86.012125,sabin2014impurities}.
By supplying such devices, the advancement of quantum technology and metrology paves the way for profound developments in many different branches of science, ranging from material sciences to biology and medicine \cite{kucsko2013nanometre,yang2011quantum}, which would otherwise be infeasible due to current less efficient and(or) invasive measurement probes.

In a generic quantum thermometer, the temperature of the bath is encoded in the evolving quantum state $\rho(t)$ of the probe and may hence be read out by measuring this state after a given $t$. 
If a large number $K$ of such independent measurements are performed, the variance $(\Delta T)^2$ of the derived temperature estimate around a rough prior estimate $T$ obeys the quantum Cramér-Rao bound \cite{Helstrom1969,PhysRevLett.72.3439,doi:10.1142/S0219749909004839,holevo2011probabilistic},
\begin{align}\label{eq:CRB}
(\Delta T)^2 \geq \frac{1}{K\mathcal{I}_T[\rho(t)]}.
\end{align}
Here $\mathcal{I}_T[\rho(t)]$ is the Quantum Fisher Information (QFI) which quantifies the information encoded in the state $\rho(t)$ at time $t$ about the temperature $T$. In an intuitive, geometric picture, it is defined by the change in the state, measured by Bures metric, as the temperature changes by an infinitesimal amount \cite{PhysRevLett.72.3439}.
There exist in general a (possibly adaptive) measurement protocol which closes the bound (\ref{eq:CRB}) as $K$ becomes large.
A well-designed thermometer should thus aim at maximizing the value of $\mathcal{I}_T[\rho(t)]$; a task which corresponds to an optimal encoding of the temperature in the state of the probe.

In conventional thermometer setups, the encoding is characterized by incoherent exchanges of energy between the probe and the bath. The temperature is thus effectively encoded in the excitation of the probe system which quickly thermalizes with the bath to reach a steady state $\rhoss$. At this point, the Fisher information saturates at the value $\mathcal{I}_T[\rhoss]$ and no further information is encoded as time progresses. Hence such a quantum thermometer operates as a classical sensor, utilizing only populations, while not including the advantages offered by quantum mechanics which rely on quantum coherences and entanglement \cite{Giovannetti19112004,PhysRevLett.96.010401,1751-8121-47-42-424006}.
Previous studies suggest that initial coherences in or simultaneous coherent driving of a single (qubit) probe system do not improve its thermometric properties; see e.g. \cite{PhysRevLett.114.220405}.
To overcome this problem, it has been proposed to map thermometry to a task of optimal phase estimation which allows quantum advantages to be utilized \cite{PhysRevA.82.011611}.

\begin{figure}
\includegraphics[trim=0 0 0 0,width=0.8\columnwidth]{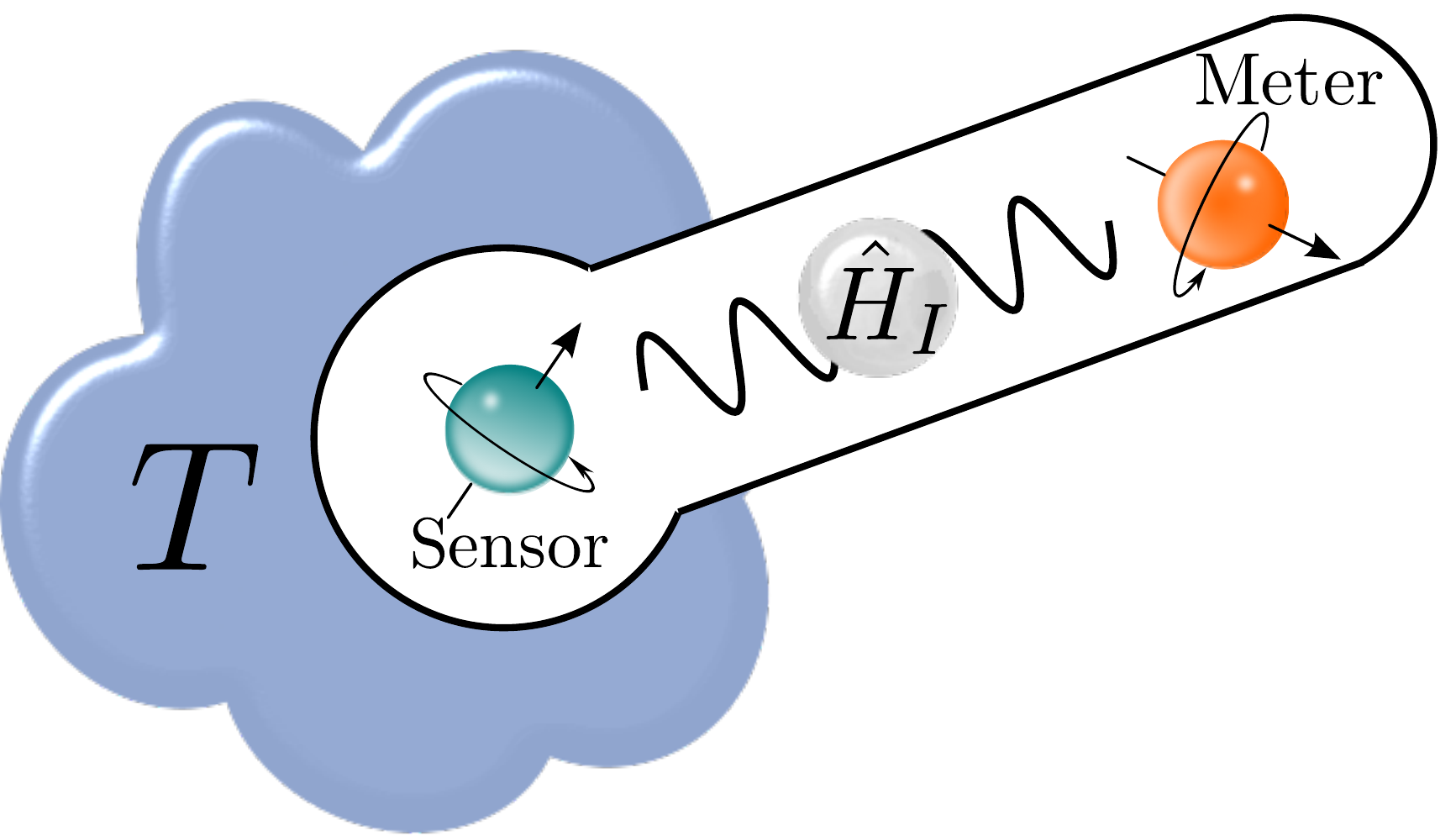}
\caption{
The temperature $T$ of a thermal bath is probed by a quantum thermometer consisting of a sensor system $\Slabel$, directly coupled to the bath, and a meter system $\Mlabel$, uncoupled from the bath but interacting via a Hamiltonian $\H_I$ with $\Slabel$.
}
\label{fig:setup}
\end{figure}

In this work, we propose a thermometer consisting of two separate quantum systems: a sensor $S$ directly coupled to the thermal bath of interest and a meter $M$ which is not directly coupled to the bath but instead serves as an information storage that can be read out at the final time $t$; see \fref{fig:setup}. While initial entanglement between the sensor and a meter system has been found to provide thermometric advantages in discriminating two distinct temperatures \cite{PhysRevA.91.012331}, we shall not rely on this effect nor on the possibility of performing joint measurements on $S$ and $M$. On the contrary, in our approach we assume the sensor and the meter to be initially uncorrelated but coupled through an interaction Hamiltonian term $\hat{H}_I$ which operates in parallel with the thermalising process affecting $S$. The main purpose of this extra dynamical contribution is to transform the sensor into an efficient information transducer between the bath and the meter. The bath-induced excitations of the sensor affect the (local) coherence terms of the meter system, creating an off-balance configuration that effectively overcomes the before-mentioned saturation problem and therby results in considerably larger values of the associated quantum Fisher information.


Our presentation is structured as follows. 
In Sec.~\ref{sec:2} we present our model and show how the dynamical evolution of the sensor-meter state may be solved analytically. 
In Sec.~\ref{sec:3} we evaluate and discuss the QFI associated with the state of the meter system, and we demonstrate the performance of our thermometer device for a two-level and a multi-level meter system.  
In Sec.~\ref{sec:LioSpec}, we discuss the advantage our proposal in terms of the structure of the Liovillian superoperator, governing the evolution of the full sensor-meter system.
Finally, in In Sec.~\ref{sec:5}, we conclude and provide an outlook.

\section{Model}\label{sec:2}
For concreteness, we assume a bosonic bath and we consider a two-level (qubit) sensor system $\Slabel$ with ground state $\ket{g}_\Slabel$ and excited state $\ket{e}_\Slabel$ whose interaction at strength $\gamma$ with the bath validates the Born-Markov approximation such that its state $\rho_\Slabel(t)$ evolves according to a master equation of the Lindblad form \cite{breuer2002theory,QMC},
\begin{align}\label{eq:ME}
\dot{\rho}_\Slabel= \mathcal{L}_T\rho_\Slabel,
\end{align}
where the Liovillian super-operator is
\begin{align}\label{eq:LT}
\mathcal{L}_T = -i\frac{\omega}{2}[\sz,\cdot]
+ \gamma_-\mathcal{D}[\hat{\sigma_-}]
+ \gamma_+\mathcal{D}[\hat{\sigma_+}],
\end{align}
and we define
$\mathcal{D}[\hat{a}] = \hat{a}\cdot\hat{a}^\dagger-\{a^\dagger a,\cdot\}/2$ with $\{\cdot,\cdot\}$ being the anti-commutator.
Here $\omega$ is the characteristic frequency of $\Slabel$ and the temperature is mapped to the evolution of the probe via the average number of resonant thermal excitations $N$, as given by the Bose distribution
\begin{align}\label{eq:bose}
N = \frac{1}{\e{\hbar\omega/ k_b T}-1},
\end{align}
causing a decay at a rate $\gamma_- = (N+1)\gamma$ and excitation at a rate $\gamma_+ = N\gamma$.

It was shown by Correa \textit{et. al.} \cite{PhysRevLett.114.220405} that an effective two-level system exhibits maximal thermal sensitivity and the use of a small quantum sensor is further motivated by the fact that often the bath is itself a nanoscale system; e.g. a micromechanical oscillator \cite{PhysRevA.84.032105}. 
Additionally, Ref.~\cite{PhysRevLett.114.220405} finds that temperature is encoded with highest accuracy in a qubit prepared in its ground state. In this case, the solution to \eqref{eq:ME} is 
\begin{align}\label{eq:rhoS}
\rho_\Slabel(t) = p_e(t)\ket{e}\bra{e}+(1-p_e(t))\ket{g}\bra{g}
\end{align}
with 
$p_e(t) = \frac{N}{2N+1}(1-\e{-(2N+1)\gamma t})$ which quickly relaxes to the Gibbs canonical ensemble at the temperature $T$ of the bath; that is $\rho_\Slabel(t)\rightarrow \rhoss_\Slabel = \frac{\ket{g}\bra{g}+\e{-\hbar{\omega}/ k_b T}\ket{e}\bra{e}}{1+\e{-\hbar{\omega}/ k_b T}}$ on time scale set by the rate $(2N+1)\gamma$.

The QFI for a measurement performed directly on the state of $S$ can be expressed as
\begin{align}\label{eq:QFIsensor}
\mathcal{I}_T[\rho_\Slabel(t)] = \frac{(dp_e/dT)^2}{p_e(1-p_e)}.
\end{align}


This function exhibits a non-monotonic behaviour \cite{PhysRevLett.114.220405,PhysRevA.96.012316} which from the zero value attained at $t=0$ brings it to
the asymptotic value  
\begin{align}\label{eq:sensorQFI}
\begin{split}
\mathcal{I}_T[\rhoss_\Slabel]
&= \left(\frac{\hbar\omega}{k_b}\right)^2\frac{\e{\hbar\omega/ k_b T}}{(1+\e{\hbar\omega/ k_b T})^2T^4}
\end{split}
\end{align}
as the state $\rho_{\Slabel}(t)$ approaches $\overline{\rho}_\Slabel$.
Although local maxima can typically be identified at finite times $t$, the global maximum of the function~(\ref{eq:QFIsensor}) corresponds to the 
maximum of (7); i.e. $\max_T\left(\mathcal{I}_T[\rhoss_\Slabel] \right)\simeq 4.53(\hbar \omega/k_b)^2$ at a temperature $k_b T\simeq 0.242\hbar\omega$. 

\subsection{Including a meter system}
While the time-independent value of the QFI, $\mathcal{I}_T[\rhoss_\Slabel]$ reflects a steady state which depends only weakly on the temperature, it is well-known in quantum metrology that the Fisher information associated with a parameter $g$ encoded in a closed quantum system by a unitary transformation $U = \e{-i g \H t}$ is given by $\I_g[\rho(t)] = 4(\braket{\H}^2-\braket{\H}^2)t^2$ \cite{PhysRevLett.96.010401,doi:10.1142/S0219749909004839}; i.e. it exhibits a persistent $t^2$-scaling with time and does not reach a constant value. 
This difference is due to the role of coherences in the latter case, and
it is enticing to seek a protocol which maps the incoherent temperature encoding in mixed state populations to coherences.

In order to circumvent the inevitable loss of coherence in the open sensor system $\Slabel$ due to the thermal coupling, we propose to achieve this goal by introducing a second \textit{meter} system $\Mlabel$ which is uncoupled from the thermal bath. The temperature is encoded in $\Mlabel$ by introducing a Hamiltonian coupling between $\Slabel$ and $\Mlabel$ of the form,
\begin{align}\label{eq:HI}
\hat{H}_I = \M\otimes\ket{e}\bra{e},
\end{align}
where $\M$ is an operator on the local space of $\Mlabel$.
If, for instance, $\Mlabel$ is a qubit, one might let $\M =\Omega/2\sx$. Beyond the simplicity of its expression, what makes such choice for $\hat{H}_I$ appealing is that it
then describes a Rabi drive of the meter qubit conditioned on $\Slabel$ being in its excited state.
Such an interaction can be realized by utilizing the dipole-dipole coupling between two spins which leads to an energy shift. For example, 
rare-earth-ion dopants in inorganic crystals have permanent electric dipole moments which are different depending on whether each ion is excited or not \cite{OHLSSON200271,PhysRevA.75.012304}.
A continuous laser illumination of a meter ion can thus be resonant when the sensor ion is in its excited state and completely off-resonant when it is in the ground state. Another well-known example is the dipole-dipole potential between neutral atoms responsible for the Rydberg Blockade mechanism \cite{PhysRevLett.85.2208}, and yet another is the hyperfine coupling between a nuclear spin and an electron spin in, e.g., NV centers \cite{childress2006coherent,PhysRevB.78.094303,PhysRevLett.109.137602}.

The state $\rho(t)$ of the full system, consisting of $\Slabel$ and $\Mlabel$ obeys a master equation
\begin{align}\label{eq:MEfull}
\dot{\rho} = -i[\hat{H_I},\rho]+\mathcal{L}_T \rho,
\end{align}
where $\mathcal{L}_T$, defined in \eqref{eq:LT}, operates locally on the sensor system.

The spectrum of $\hat{H}_I$ can be seen as a sequence of effective two-level systems, uncoupled by the thermal interaction ($\mathcal{L}_T$), with ground states $\ket{m}\otimes\ket{g}$ and excited states $\ket{m}\otimes\ket{e}$ where the $\ket{m}$ are eigenstates of the operator $\M$ with corresponding eigenvalues $\lambda_m$,
\begin{align}
\M\ket{m} = \lambda_m\ket{m}.
\end{align}
The total population difference between the upper $\{\ket{m}\otimes\ket{e}\}_m$ and lower manifolds $\{\ket{m}\otimes\ket{g}\}_m$ hence represents the information available from $\Slabel$ alone, while 
the information encoded in $\Mlabel$ is represented by the coherences amongst the individual two-level transitions.

Following this idea, we expand $\rho(t)$ in the eigenbasis of $\M$,
\begin{align}
\rho(t) = \sum_{m,m'} A_{mm'}\ket{m}\bra{m'}\otimes \rho_{mm'}(t).
\end{align}
Here the $\rho_{mm'}(t)$ operate on the sensor qubit space, and the 
$A_{mm'} = \bra{m'}\rho_\Mlabel(t=0)\ket{m}$ are defined by the initial state
$\rho_\Mlabel(t=0)$ of $\Mlabel$.
From the master equation (\ref{eq:MEfull}), the equations of motion for the $\rho_{mm'}(t)$ are seen to be,
\begin{align}
\begin{split}
\dot{\rho}_{mm'} = &-i \frac{\lambda_m+\lambda_{m'}}{2}\left[\ket{e}\bra{e},\rho_{mm'}\right]
\\
&-i\frac{\Omega_{mm'}}{2}\left\{\ket{e}\bra{e},\rho_{mm'}\right\}+\mathcal{L}\rho_{mm'},
\end{split}
\end{align}
where $\Omega_{mm'}= \lambda_m-\lambda_{m'}$. 
The commutator term does not have any effect for the case of a sensor initialized in the ground state as assumed here. The diagonal elements with $\Omega_{mm} = 0$ hence solve \eqref{eq:ME}; i.e. one finds
$\rho_{mm}(t) = \rho_\Slabel(t)$ as given in \eqref{eq:rhoS}.
The anti-commutator term is not trace preserving, and the solutions for the coherences,
\begin{align}\label{eq:rhommp}
\begin{split}
\rho_{m m'}(t) = &\frac{\e{-[\gamma(N+1/2)+i\Omega_{mm'}/2]t}}{\alpha}
\Big(\gamma N [\e{\alpha t /2}-\e{-\alpha t /2}]\ket{e}\bra{e}
\\
&+
\frac{1}{2}[(\gamma+i\Omega_{m m'})(\e{\alpha t /2}-\e{-\alpha t /2})
\\
&+
\alpha(\e{\alpha t /2}+\e{-\alpha t /2})]\ket{g}\bra{g}
\Big)
,
\end{split}
\end{align}
with 
$\alpha(N) = \sqrt{(2N+1)\gamma^2-\Omega_{mm'}^2+2i\gamma\Omega_{mm'}}$,
a complex parameter, are not normalized but rather decay to zero at long times. 

\section{Quantum Fisher information}\label{sec:3}
As detailed above, the simple form of the Hamiltonian \eqref{eq:HI} allows the dynamical evolution of the full system to be solved analytically for a general meter operator $\M$, and it is clear that the solution and hence the thermometric properties of our device depend only on the spectrum of the operator $\M$. 

Tracing out $\Mlabel$, we recover the thermalizing state of $\Slabel$, $\mathrm{Tr}_\Mlabel(\rho) = \sum_m A_{mm} \rho_ {mm}(t) =\rho_\Slabel(t)$
where we used that $\sum_m A_{mm} = 1$. Note that in the partial trace operation all coherence terms with $m\neq m'$ cancel.
This implies that the QFI associated with a measurement on the sensor $\Slabel$ alone is not influenced by the presence of the meter $\Mlabel$ and is indeed encoded in the total population difference between the manifolds as argued above. 

The reduced state of $\Mlabel$ is given by 
\begin{align}\label{eq:meterState}
\begin{split}
\rho_\Mlabel(t) 
&= \sum_{m}A_{mm}\ket{m}\bra{m}+ \sum_{m\neq m'} A_{mm'} \mathrm{Tr}_\Slabel\left(\rho_{mm'}\right)\ket{m}\bra{m'},
\end{split}
\end{align}
where we used that the $\rho_{mm}(t)$ obey a trace preserving master equation. 
Since the first sum depend only on the initial state of $\Mlabel$, it is evident that the temperature is indeed encoded purely in its coherences.
Furthermore, it is clear that the performance depends critically on the initial preparation of $\Mlabel$. If, for instance, it is prepared in an eigenstate $\ket{n}$, we have $A_{m m'} = \delta_{m' n}\delta_{n m}$, and its state $\rho_\Mlabel(t)$ is temperature independent. 
The optimal initial state $\ket{\psi_\Mlabel(t=0)}=\sum_m c_m\ket{m}$, which due to the convexity on the QFI is pure, depends in general on the spectrum of the operator $\M$, but we note that since any phases correspond to a unitary transformation of the meter state, to which the QFI is invariant \cite{doi:10.1142/S0219749909004839}, the $c_m$ can be taken as real and positive.

\subsection{Example: Two-level meter} \label{sec:twoLevel}
Our main example concerns a meter system with two levels, $\ket{0}$ and $\ket{1}$, and for concreteness we shall let $\M = \Omega \sx /2$, corresponding to a conditional Rabi drive of $\Mlabel$ as explained above. To maximize the coherences in the eigenbasis of $\M$, the meter should be prepared in a state $\ket{0} = (\ket{+}+\ket{-})/\sqrt{2}$, where $\ket{\pm}$ are the eigenstates of $\sx$.

\begin{figure}
\includegraphics[trim=0 0 0 0,width=0.9\columnwidth,left]{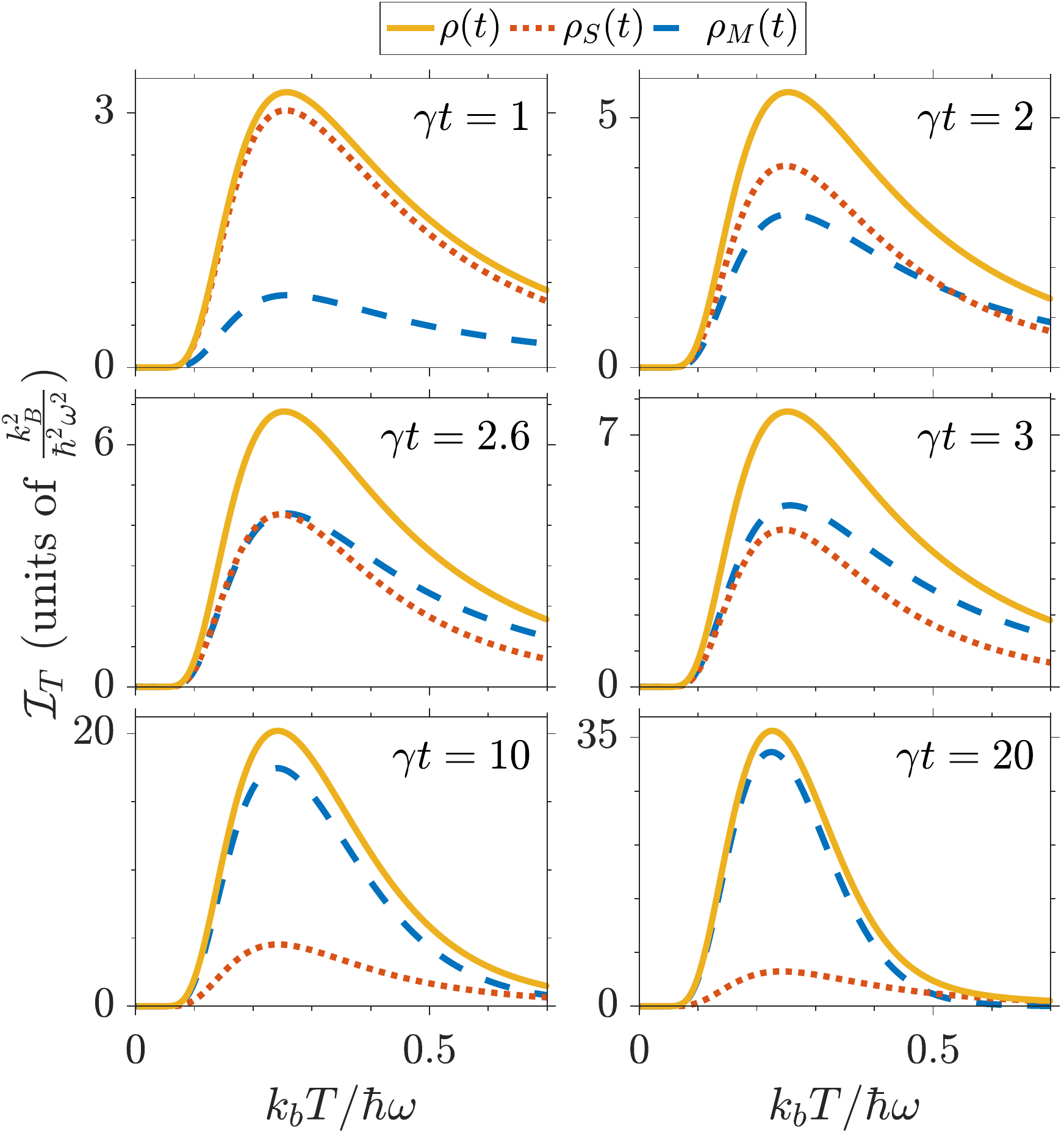}
\caption{Quantum Fisher information $\I_T[\cdot]$ associated with the full state $\rho(t)$, the sensor state $\rho_\Slabel(t)$ or the meter state $\rho_\Mlabel(t)$.
Results are shown as a function of the temperature $T$ and for $\Omega=2\gamma$. The panels correspond to different probing times $t$ as annotated in the figure window.
}
\label{fig:compareQFI}
\end{figure}

In \fref{fig:compareQFI}, we compare the QFI associated with  either of the reduced states, $\rho_\Slabel(t)$ or $\rho_\Mlabel(t)$, to that of the full sensor-meter state $\rho(t)$. Results are shown as a function of the temperature $T$ and for different probing times in each panel. At short times, $\gamma t = 1$, the thermometric information is held mainly by $\Slabel$ ($\I_T[\rho_\Slabel(t)]\simeq \I_T[\rho(t)]$) but as time progresses, temperature dependent coherences build up in $\Mlabel$ and while $\Slabel$ reaches a steady state with maximum information (\ref{eq:sensorQFI}), the information in the meter $\Mlabel$ keeps increasing. Hence, at $\gamma t\simeq 2.6$ we have ($\I_T[\rho_\Mlabel(t)]\simeq \I_T[\rho_\Slabel(t)]$), and at larger times $\gamma t = 20$ the information in the combined state is held predominantly by $\Mlabel$. Furthermore, at this point $\I_T[\rho(t)]\simeq \I_T[\rho_\Mlabel(t)] \gg 4.53 (\hbar \omega/k_B)^2\geq \I_T[\rho_\Slabel(t)]$. Evidently, the capability of the meter system to accumulate information for a much longer time allows it to reach a significantly larger thermometric sensitivity.

It is an attractive feature of our device that after some initial time, a local measurement on the meter $\Mlabel$ is able to extract almost all the information from the state. This makes the thermometer more feasible to implement, and at the same time less invasive since $\Mlabel$ may, as depicted in \fref{fig:setup}, be located outside, e.g., a biological sample. 
To characterize our quantum thermometer, we shall thus focus on the long-time behavior of the QFI, associated with the reduced state $\rho_\Mlabel(t)$ of the meter $\Mlabel$ alone. 

\begin{figure}
\subfloat[\label{fig:twoLevelQFIcolor}]{
\includegraphics[trim=0 0 0 0,width=0.98\columnwidth]{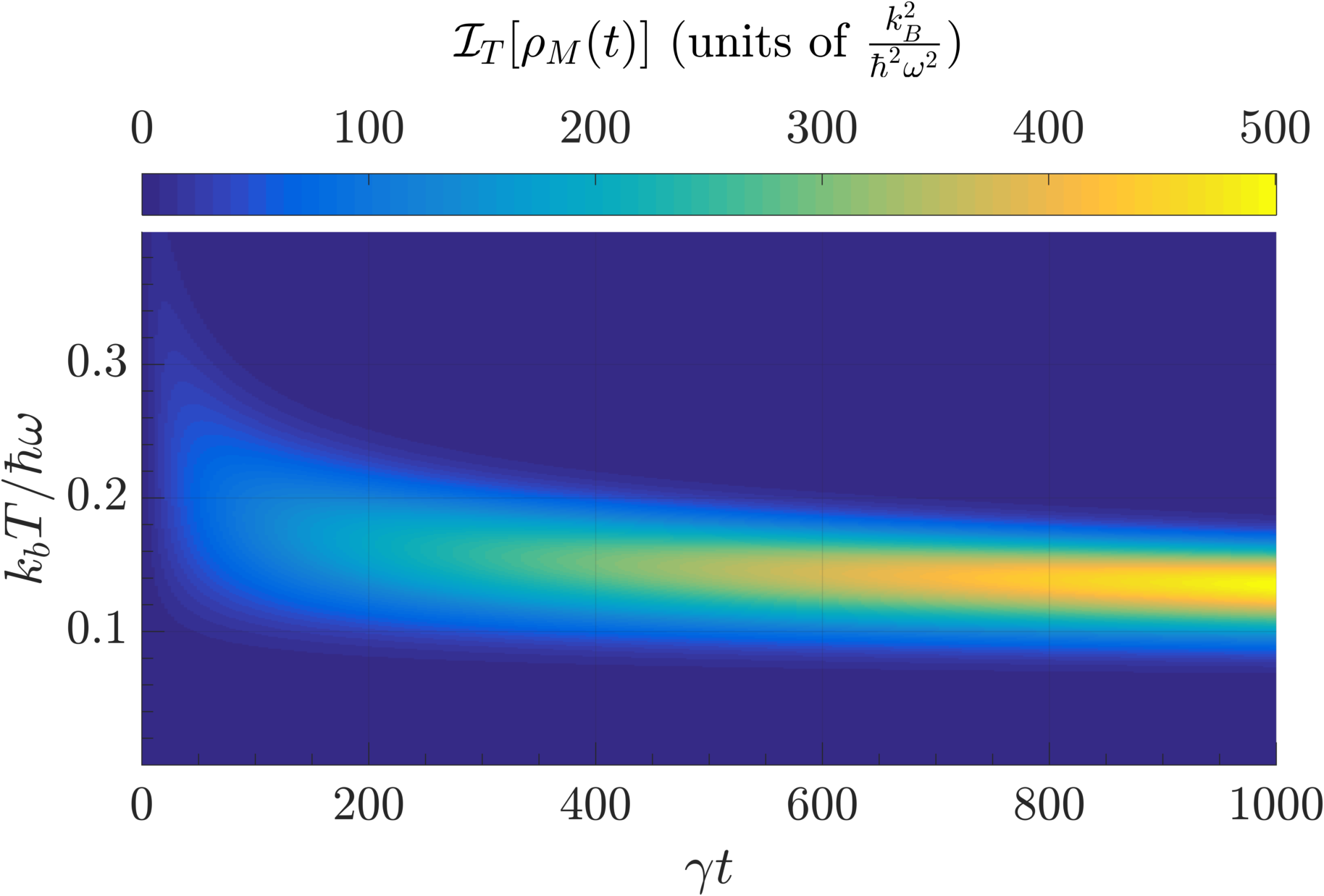}
\centering}

\subfloat[\label{fig:twoLevelQFIsnaps}]{
\includegraphics[trim=0 0 0 0,width=0.98\columnwidth]{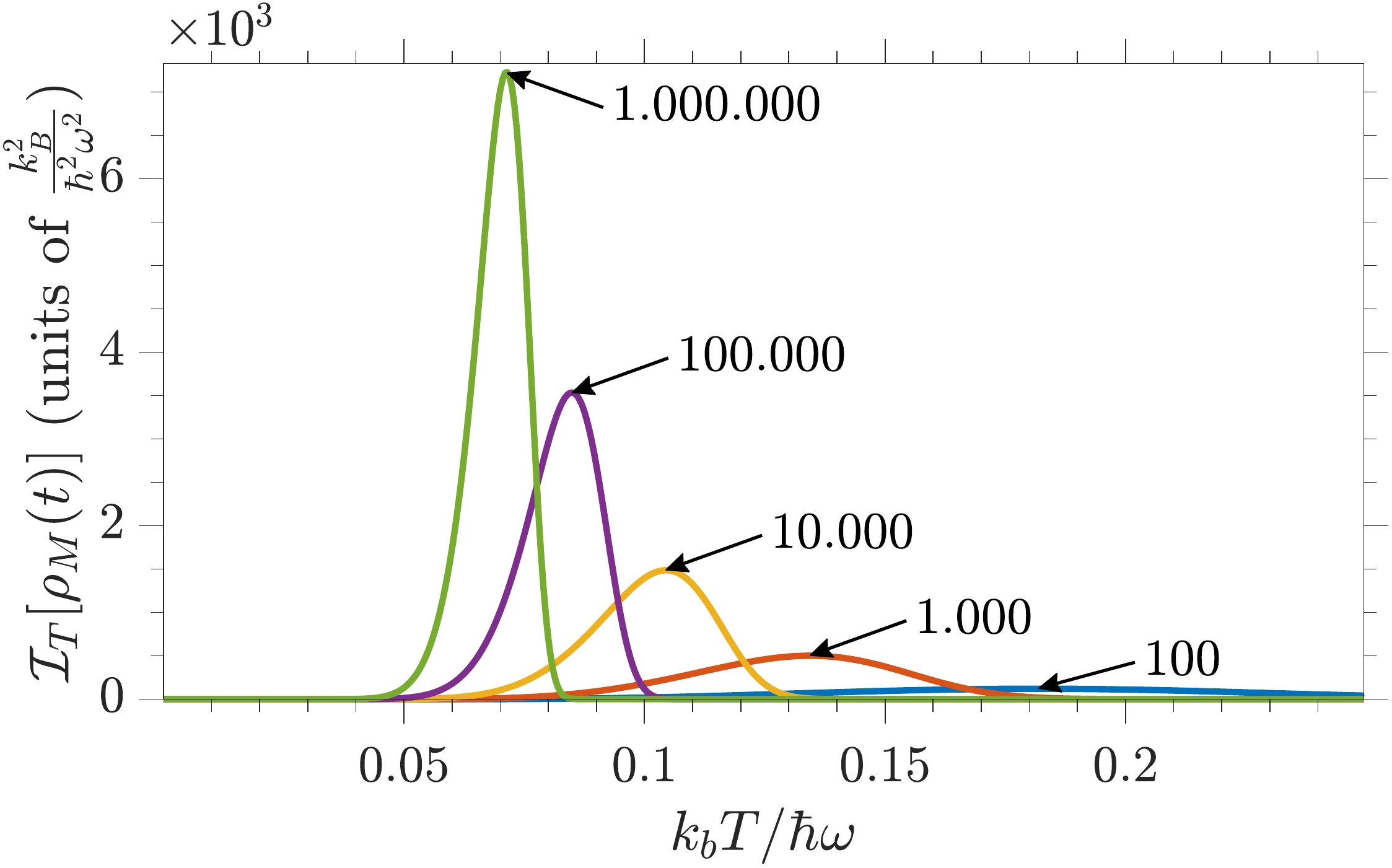}
\centering}
\caption{Quantum Fisher information $\mathcal{I}_T[\rho_\Mlabel(t)]$ associated with estimating the temperature $T$ from a local measurement on a two-level meter system $\Mlabel$ coupled to the sensor $\Slabel$ via a Hamiltonian $\H_I = \Omega/2\sx\otimes \ket{e}\bra{e}$ with $\Omega = 2\gamma$.
(a) Color plot showing the dependence of $\mathcal{I}_T[\rho_\Mlabel(t)]$ on the temperature $T$ and the probing time $t$. 
(b) Curves for $\mathcal{I}_T[\rho_\Mlabel(t)]$ as a function of $T$ are shown for different probing times $\gamma t=100,1.000,10.000,100.000,1.000.000$ as annotated with arrows in the figure window.
}
\label{fig:twoLevel}
\end{figure}

\begin{figure}
\includegraphics[trim=0 0 0 0,width=0.98\columnwidth]{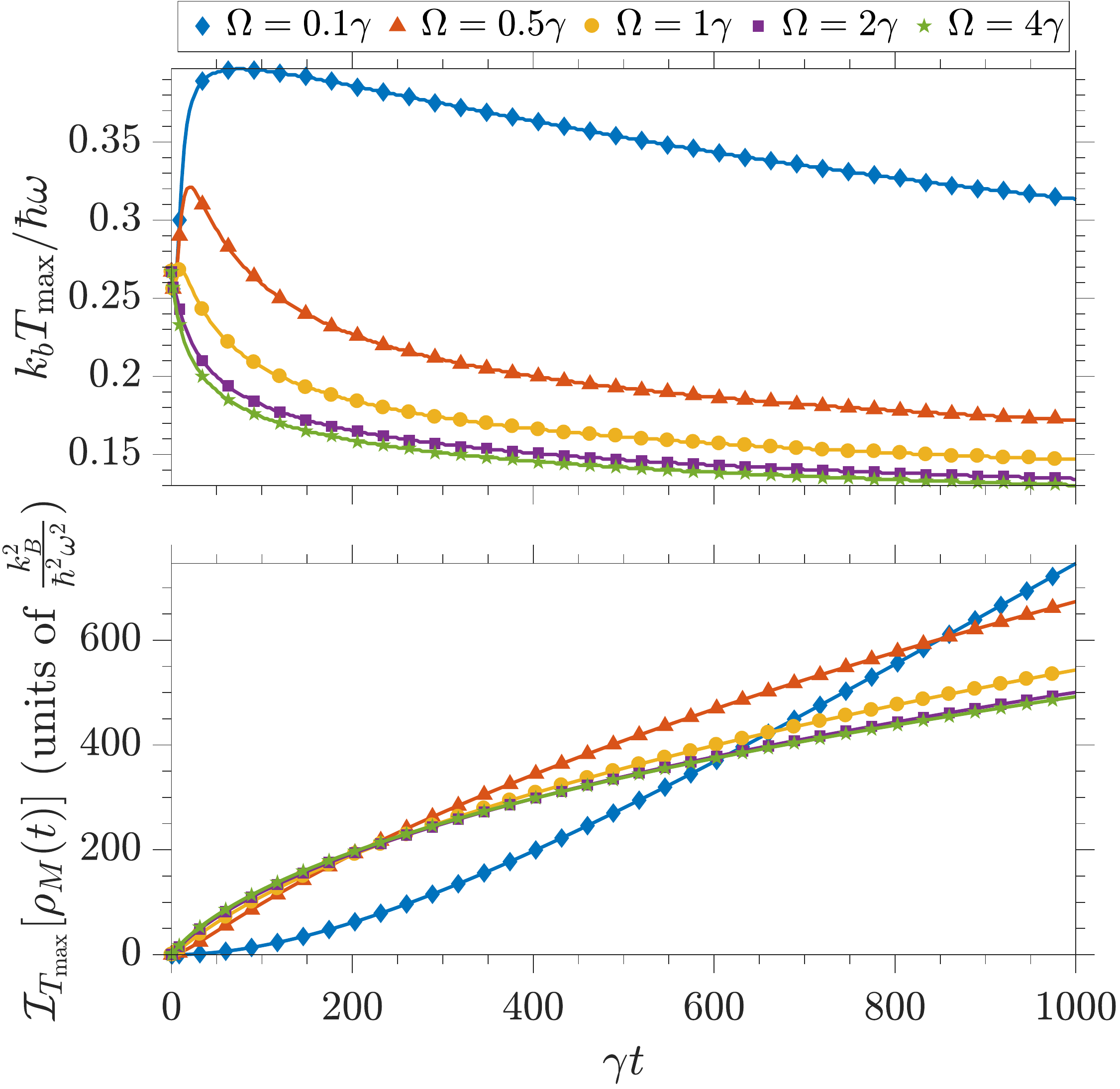}
\centering
\caption{
The temperature $T_{\mathrm{max}}$ (upper panel) at which the QFI takes its maximum value $\mathcal{I}_{\Tm}[\rho_\Mlabel(t)]$ (lower panel) as a function of the probing time. 
Results are shown for a two-level meter system $\Mlabel$ and with $\H_I = \Omega/2\sx\otimes \ket{e}\bra{e}$ for different values of $\Omega$.
}
\label{fig:twoLevelTmax}
\end{figure}

The QFI of a two-level density matrix may be expressed as \cite{dittmann1993riemannian},
\begin{align}\label{eq:twoLevelQFI}
\I_T[\rho_\Mlabel] = 4\mathrm{Tr}\left[\rho_\Mlabel\left(\frac{\partial \rho_\Mlabel}{\partial T}\right)^2\right] + \frac{1}{\det(\rho_\Mlabel)}\left[\frac{\partial \det(\rho_\Mlabel)}{\partial T}\right]^2,
\end{align}
and the color plot in \fref{fig:twoLevel}(a) shows an example of its evolution from time $\gamma t=0$ to $\gamma t=1.000$ for a relevant range of temperatures.
In 3(b) we plot $\I_T[\rho_\Mlabel(t)]$ at specific times from $t=100\gamma^{-1}$ to $t=100.000\gamma^{-1}$. 
The sensitivity of $\Mlabel$ depends on the temperature $T$ relative to the frequency $\omega$ of $\Slabel$ and reaches a maximum at a temperature $\Tm(t)$ which, as seen in \fref{fig:twoLevelTmax}, decreases with time (we will come back to this point). The QFI, $\mathcal{I}_{T}[\rho_\Mlabel(t)]$ at and around this temperature reaches values which are much larger than the sensitivity offered by a sensor qubit alone; $\mathcal{I}_T[\rho_\Slabel(t)] \leq 4.53217(\hbar \omega/k_b)^2$.
It should be noted that \fref{fig:twoLevelQFIsnaps} shows how the temperature-range, at which $\Mlabel$ is sensitive, decreases with time, and that though it appears that the sensitivity at $\Tm(t)$ increases without bounds, our treatment of the thermal coupling in the Born-Markov approximation breaks down for very low temperatures where strong correlations between the sensor $\Slabel$ and the bath may appear.

The strength $\Omega$ of the Hamiltonian (\ref{eq:HI}) appears as a control parameter, and in \fref{fig:twoLevelTmax}, we show $\Tm(t)$ and the corresponding QFI, $\mathcal{I}_{\Tm}[\rho_\Mlabel(t)]$ for different values of $\Omega$.
For small $\Omega\lesssim \gamma$, the sensitivity of $\Mlabel$ can be tuned within a relatively broad interval by adjusting $\Omega$, while it is near independent for stronger interactions; see $\Omega=2,4\gamma$ in the figure.
The value of $\mathcal{I}_{\Tm}[\rho_\Mlabel(t)]$ is at short times larger for strong couplings while at later times it is favorable to apply a weaker laser field to the meter $\Mlabel$. 
This can be understood by a competition between the two roles played by $\H_I$: i) to transfer information about the temperature from $\Slabel$ to $\Mlabel$, and ii) to mediate decoherence between the two systems. Hence, at short times it is favorable to transfer a large amount of information quickly at the cost of a faster dephasing of that information, while when longer time is available a slower transfer is compensated by a longer coherence time of $\Mlabel$.  

From Figs.~(\ref{fig:compareQFI},\ref{fig:twoLevel}), it is clear that the interesting regime concerns large times, and that the relevant temperature range is centered around $k_bT/\hbar\omega \simeq 0.2$ which corresponds to small values $N\simeq 0.007$ of the thermal bath excitation \cite{de2016local,PhysRevLett.114.220405}. Assuming then $\Omega \gg \gamma N$ and $\gamma t \gg 1$, we obtain a simple approximation for the QFI,
\begin{align} \label{eq:longTimeQFI}
\begin{split}
\mathcal{I}_T[\rho_\Mlabel(t)]&\simeq  \left(\frac{dN}{dT}\right)^2 \frac{\gamma^2t^2\e{-2\Gamma_N t}}{\Omega^2+\gamma^2}
\\
&\times
\left(\Omega^2+4\gamma^2 N^2+\frac{(\Omega^2-2\gamma^2N)^2}{(\Omega^2+\gamma^2)(\e{2\Gamma_N t}-1)}\right),
\end{split}
\end{align}
where $dN/dT$ is the differential of the Bose distribution (\ref{eq:bose}) with respect to temperature and we point out that the effective decay rate $\Gamma_N = N\gamma(\Omega^2-N\gamma^2)/(\Omega^2+\gamma^2)$ is very small.
We thus see that the QFI scales as $\propto\gamma^2t^2\e{-2\Gamma_N t}$, and for any given temperature ($N$) it reaches a maximum value at the time, $t_{\mathrm{max}}(T) = \Gamma_N^{-1}$ after which it decreases to zero as the coherences (\ref{eq:rhommp}) decay. This time, however, appears later for smaller values of $N$ leading to the decrease in time of $\Tm(t)$ seen in \fref{fig:twoLevelTmax}. Still, we want to stress that for any temperature $T$, the QFI is upper bounded by $\mathcal{I}_T[\rho_\Mlabel(t_{\mathrm{max}}(T))]$, and that at very large times $t\gg t_{\mathrm{max}}(T)$ the coherences $\mathrm{Tr}_\Slabel\left(\rho_{\pm\mp}\right)$ in $\Mlabel$ vanish such that, according to \eqref{eq:meterState}, it is left in a statistical mixture $\rho_\Mlabel(t) =\mathbb{I}/2$  with no information regarding the temperature of the bath.

%

\subsection{A multi-level meter system}
For a meter system $\Mlabel$ of arbitrary dimension $n$, the quantum Fisher information 
constitutes a complicated expression even when an analytic expression is known for the mixed state density matrix of the system. Here we treat these cases . For this purpose we apply the following equivalent form of the QFI
\cite{PhysRevA.97.042322},
\begin{align}\label{eq:QFI}
I_T[\rho_\Mlabel] = 2 \partial_T\vv{\rho_\Mlabel}^\dagger \left(\rho^*_\Mlabel\otimes\mathbb{I}+\mathbb{I}\otimes \rho_\Mlabel\right)^{-1}\partial_T\vv{\rho_\Mlabel},
\end{align}
where $\vv{\rho}$ denotes vectorization of the density matrix $\rho$.

To generalize the two-level example studied in Sec.~\ref{sec:twoLevel} we focus on a meter operator $\M  = \Omega \hat{S}_x$, where $\hat{S}_x$ is the x-component of the spin in a spin-$(n-1)/2$ system.  
If $\Mlabel$ is composed of several qubits, we have $\hat{S}_x = \sum \hat{\sigma}_x^{(i)}$ where $\hat{\sigma}_x^{(i)}$ operates on qubit $i$.
\begin{figure}
\includegraphics[trim=0 0 0 0,width=0.98\columnwidth]{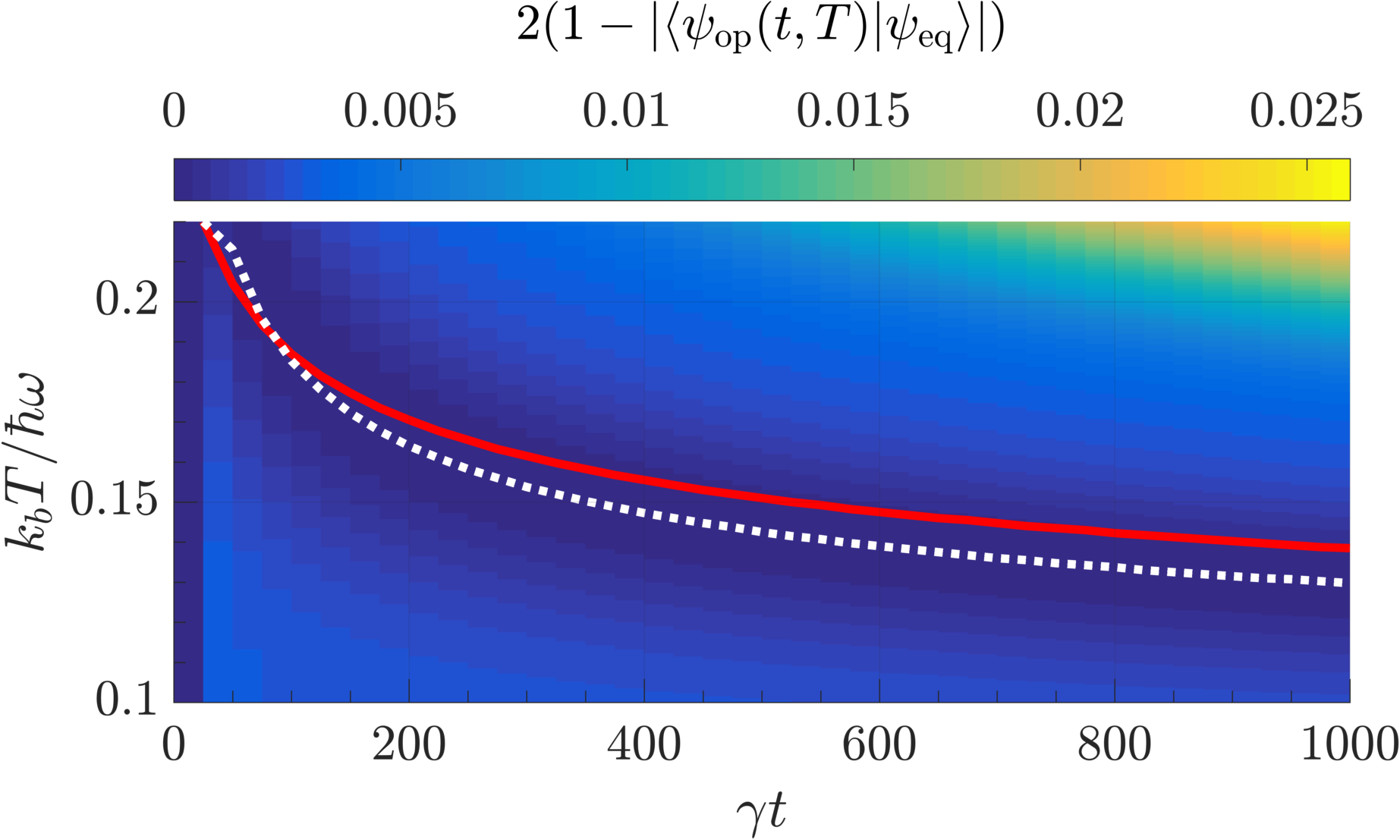}
\centering
\caption{Color plot depicting for a range of temperatures $T$ and total probing times $t$, the Bures distance, $2(1-|\langle\psi_{\mathrm{op}}(t,T)|\psi_{\mathrm{eq}}\rangle|)$ from an equal superposition $\ket{\psieq} = \frac{1}{\sqrt{n}}\sum \ket{m}$ of the eigenstates of the operator $\M$ to the initial state $\ket{\psiop(t,T)}$ of $\Mlabel$, which maximizes the value of $\I_T[\rho_\Mlabel(t)]$.
The dotted, white line tracks the temperature where $\ket{\psiop(t,T)} = \ket{\psieq}$.
The full, red line tracks the temperature $\Tm(t)$ for which $\I_T[\rho_\Mlabel(t)]$, evaluated with initial meter state $\ket{\psi_\Mlabel} = \ket{\psiop(t,T)}$, is maximal.
Results are shown for a meter system with $n=6$ levels. 
}
\label{fig:initialState}
\end{figure}

While identification of the optimal initial state as an equal super position $\ket{\psieq} = \frac{1}{\sqrt{n}}\sum_m \ket{m}$ of the eigenstates of $\M$ was straightforward in the two-level case, the general case is more complicated. Rather than just maximizing the initial coherences, the different values of the $\Omega_{mm'}$ must be taken into account, and in general we have recourse to numerical maximization of the QFI over all possible initial configurations with positive coefficients $c_m$ 
The optimal state $\ket{ \psiop(t,T)}$ depends on both the probing time $t$ and the temperature $T$. In \fref{fig:initialState} we plot the Bures distance of $\ket{\psiop(t,T)}$ from $\ket{\psieq}$ as a function of $t$ and $T$ and for $n=6$. It is seen that while $\ket{\psiop(t,T)}$ is in general different from $\ket{\psieq}$, the discrepancy is moderate and for a given time-dependent temperature (dotted, white curve) it vanishes. This temperature is close but not equal to $\Tm(t)$ as tracked by the red line. We find that the distance from an equal super position shows a similar functional dependence on $t$ and $T$ for other values of $n$.

In any real thermometry task, the precise temperature is unknown so rather than defining an initial state $ \ket{\psi_\Mlabel(t=0)}$ which depends on the specific value of $T$, one has recourse to select a specific state regardless of the precise temperature. The results in \fref{fig:initialState} and the intuition regarding the role of coherences suggest that in general one can expect near optimal results by setting $\ket{\psi_\Mlabel(t=0)}= \ket{\psieq}$ which we shall assume in the remainder of this section.

\begin{figure}
\includegraphics[trim=0 0 0 0,width=0.98\columnwidth]{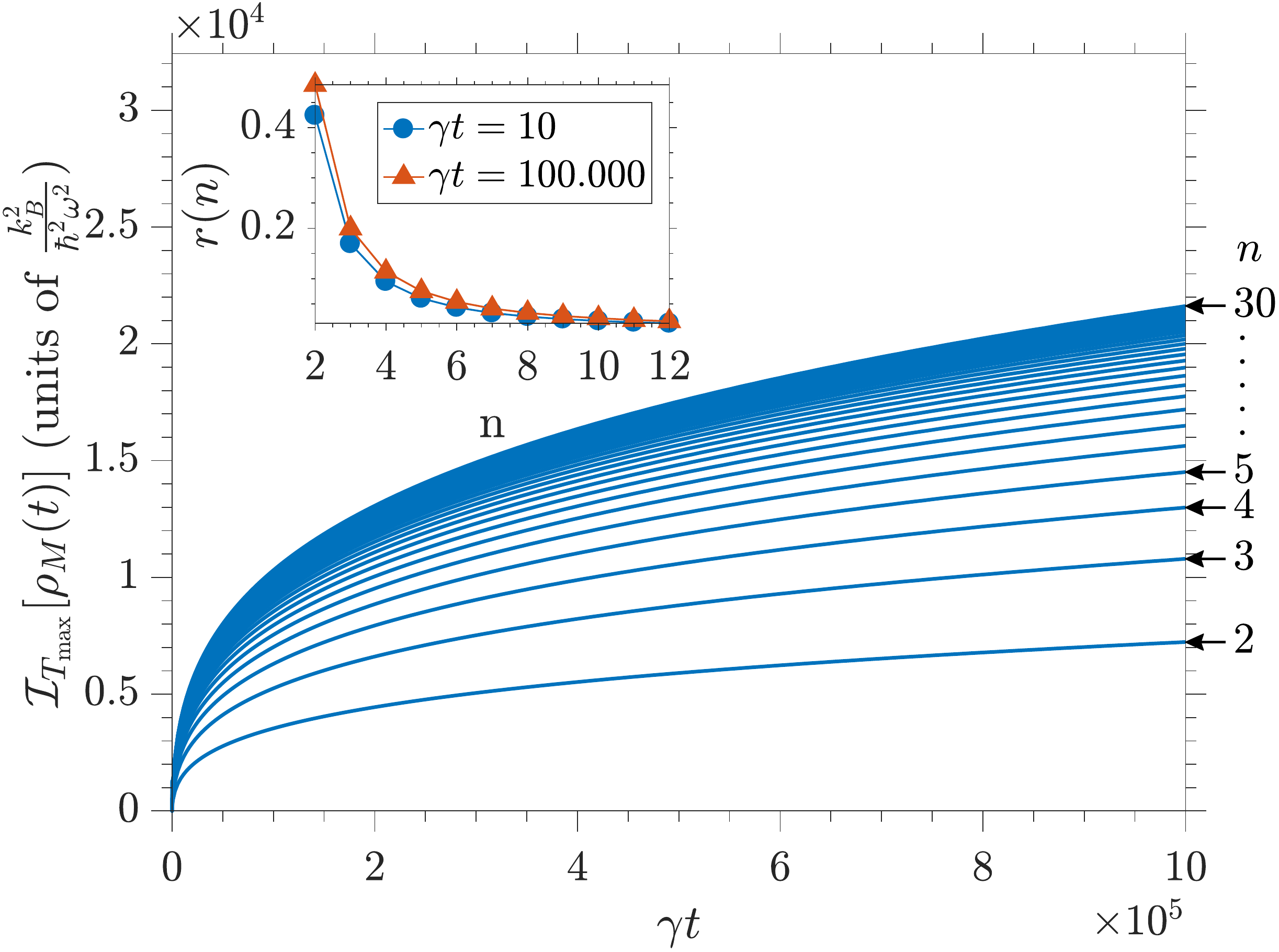}
\centering
\caption{ 
Time evolution of the quantum Fisher information $\mathcal{I}_{\Tm}[\rho_\Mlabel(t)]$ associated with estimating the temperature $\Tm(t)$ which maximizes its value from a local measurement on an $n$-level meter system coupled to $\Slabel$ via a Hamiltonian $\H_I = \Omega\hat{S}_x\otimes \ket{e}\bra{e}$ with $\Omega = 2\gamma$.
Results are shown for $n=2,3,...,30$. The inset depicts the relative scaling,
$r(n) = (\mathcal{I}^{(n+1)}_{\Tm}-\mathcal{I}^{(n)}_{\Tm})/\mathcal{I}^{(n)}_{\Tm}$, of $\mathcal{I}^{(n)}_{\Tm}$ with $n$ at short times $\gamma t=10$ and at long times $\gamma t = 100.000$.
}
\label{fig:nDependence}
\end{figure}

We proceed to probe the advantage of adding more levels to the meter system $\Mlabel$. We find that $\I_T[\rho_\Mlabel(t)]$ is maximized around the same time dependent value $\Tm(t)$ (see \fref{fig:twoLevelTmax}(a)) independently of $n$, and in \fref{fig:nDependence} we show the QFI at this temperature as a function of time and for different values of $n$.
It is seen that for all times, $\mathcal{I}_{\Tm}[\rho_\Mlabel(t)]$ increases with $n$, signifying that a higher dimensional meter system allows a larger sensitivity to small temperature variations.
This result suggests that the optimal meter system is a harmonic oscillator with an infinite dimensional Hilbert space.
Notice, however, that the gain saturates for larger $n$ as the curves are seen to lie closer and closer. In the inset we quantify this by showing how the relative increase in the QFI as one more level is added to $\Mlabel$ is (near) time-independent and approaches zero for $n\gtrsim 10$.

\section{Temperature dependence in the Liovillian spectrum} \label{sec:LioSpec}
By discussing the emergence of coherences, we have provided an intuitive understanding of the advantage offered by coupling the sensor $\Slabel$ to a meter system $\Mlabel$. In this section we explain how this advantage can be understood from the structure of the Liovillian superoperator $\mathcal{L} = -i[\H_I,\cdot]+\mathcal{L}_T$, governing, via. \eqref{eq:MEfull}, the encoding of the temperature in the full sensor-meter state.


In the long time limit, this super-operator will asymptotically bring the joint sensor-meter state to the
stationary eigenspace associated with its null eigenvalue.
The convergence of this process is exponential and determined by the inverse of the smallest modulus of the real parts of its non-zero eigenvalues. 
(which by construction are all non-positive). Accordingly, in this regime we can write 
\begin{align}
\rho(t) \simeq \Pi_0 \rho(0) + \sum_j \e{\lambda_j t} \Delta_j, 
\end{align}
where $\Pi_0$ is the projector on the null eigenspace of $\mathcal{L}$ and the summation involves those non-zero eigenvalues $\lambda_j$ of $\mathcal{L}$ that 
have the smallest (in modulus) real component with corresponding state components $\Delta_j$. 

The quantum Fisher information (\ref{eq:twoLevelQFI},\ref{eq:QFI}) refers to the derivative of the state at time $t$ with respect to the
temperature, i.e. to
\begin{align}\label{eq:rhoderivative}
\partial_T \rho(t) = ( \partial_T \Pi_0 ) \rho(0) +  \sum_j \e{\lambda_j t} (t\partial_T  \lambda_j  \Delta_j + \partial_T   \Delta_j).
\end{align}
As $t$ diverges only the first term survives and the related QFI derives from
\begin{align}
\partial_T\rho(t) \simeq  (\partial_T \Pi_{0})\rho(0).
\end{align}
Hence no scaling with time remains, and the QFI is given by that of $\Slabel$ alone, \eqref{eq:sensorQFI}.

If, however, some eigenvalues indexed by $l$ have real parts very close to zero, the contribution from their part of the spectrum in \eqref{eq:rhoderivative} persist for very long and one may indeed see terms in the QFI (\ref{eq:QFI}) scaling as $\propto t^2$ until times $t\gg \mathrm{Re}(\lambda_l)^{-1}$.
In other words, allowing the temperature to be encoded in the eigenvalues and not just the projectors may provide a significant metrological advantage.

\begin{figure}
\includegraphics[trim=0 0 0 0,width=0.98\columnwidth]{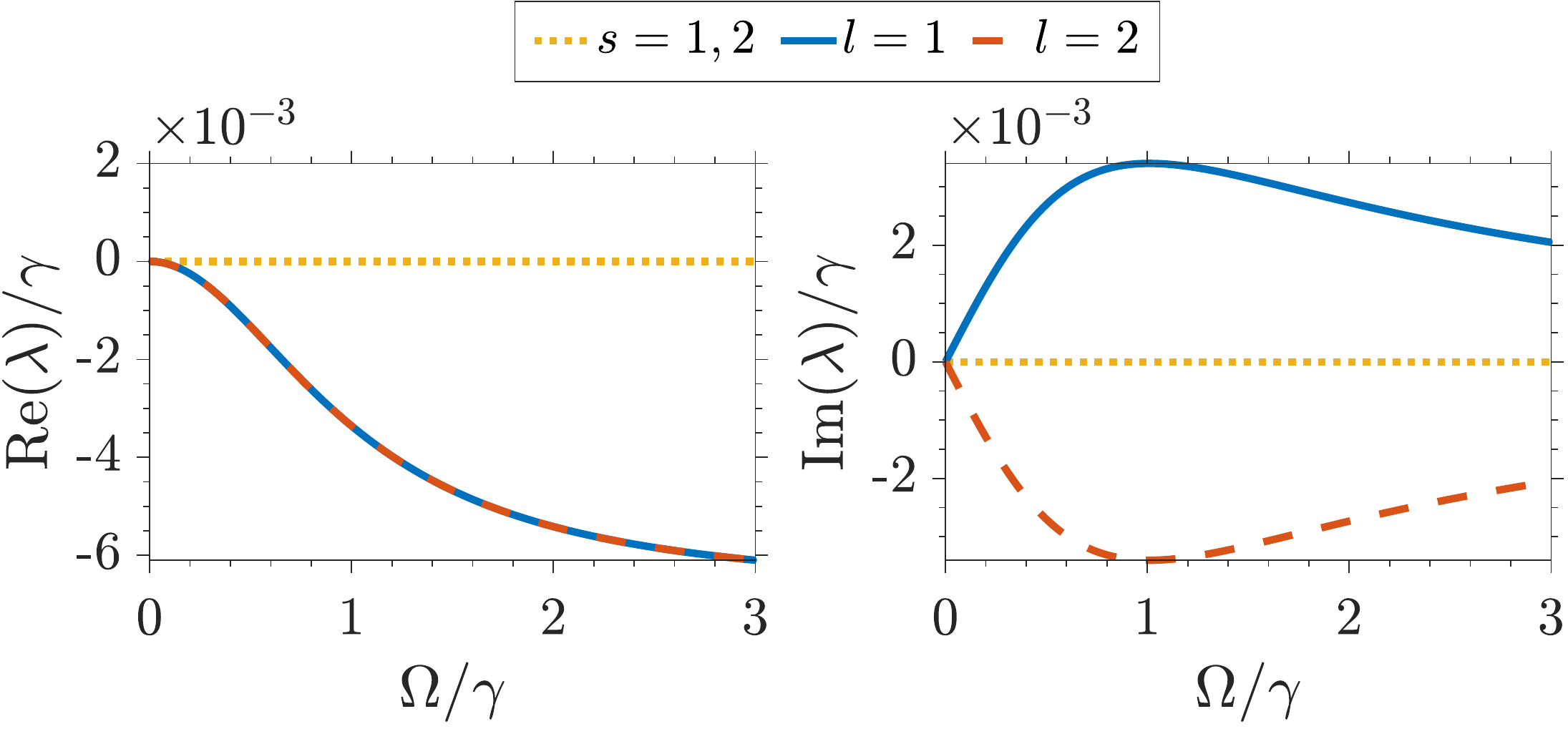}
\centering
\caption{Real (left panel) and imaginary (right panel) parts of the four eigenvalues  with the largest real parts of the superoperator $\mathcal{L}$ governing the evolution of the full sensor-meter system Results are shown as a function of the strength of the interaction Hamiltonian $\Omega$ and for a temperature $k_B T=0.2\hbar \omega$.
}
\label{fig:eigs}
\end{figure}

This is exactly the case for our thermometer device. For the two-level example of Sec.~\ref{sec:twoLevel}, we show in \fref{fig:eigs} the real and imaginary parts of the four eigenvalues with largest real parts as a function $\Omega$ for a temperature $k_B T=0.2\hbar \omega$.
Without a meter ($\Omega=0$), the four eigenvalues are all zero. The effect of adding a meter ($\Omega>0$) is to lift the degeneracy of two of the eigenvalues 
\begin{align}
\begin{split}
\lambda_{l=1} &= \frac{1}{2}\left[-(2N+1)\gamma +i \Omega -\alpha^*(N)\right]
\\
\lambda_{l=2} &= \frac{1}{2}\left[-(2N+1)\gamma -i \Omega +\alpha(N)
\right],
\end{split}
\end{align}
while the other two ($s=1,2$), corresponding to the steady state, remain zero. The very small value of $|\mathrm{Re}(\lambda_l)|$ combined with non-zero, temperature $(N)$ dependent values of $\mathrm{Im}(\lambda_l)$ are crucial for the success of our thermometer. 
\\


\section{Conclusion and outlook}   
\label{sec:5}
We have proposed a quantum thermometer which maps the incoherent encoding of a temperature from a sensor system to coherences in a meter system by a realistic Hamiltonian interaction. 
The coherent encoding allows the meter state to exhibit a much larger temperature-sensitivity than the sensor state alone. 
While an effective two-level system has been identified as an optimal temperature sensor \cite{PhysRevLett.114.220405}, we find that the sensitivity increases with the dimensionality of the the meter system.
For simplicity we focused on bosonic bath in our presentation but calculations show that similar results are valid in the case of a fermionic reservoir.

From our examples, it is clear that the achievements and sensitivity range of our thermometer device depends in a complicated manner on the strength and the spectrum of the interaction Hamiltonian (\ref{eq:HI}), on the initial preparation of the meter system, and on the total time $t$ available in a given experiment. 
Hence, an appropriate meter should be designed for the specific task at hand, e.g taking into account the expected temperature and experimental constraints.
   
In a broader context, our protocol effectively increases the sensitivity to an unknown parameter $g$ by interfering with the encoding of that parameter in the state of a quantum system.
Such possibilities are highly relevant but yet fairly unexplored in quantum metrology. 
It has been proven that the QFI associated with an unitary encoding by a Hamiltonian of the form $\H(g) = g \H_0$  can not be enhanced by adding a second $g$-independent Hamiltonian term \cite{PhysRevA.88.052117}.
However, the protocol presented in the current work constitutes an example where
the sensitivity to a parameter (temperature), encoded by an incoherent interaction with a bath, is in fact improved by adding a suitable Hamiltonian interaction.
It would be interesting to investigate more generally under which circumstances and how such an improvement is possible when parameters are encoded in an open system by a Liovillian operator.
Such an analysis could be guided by the ideas presented in Sec.~\ref{sec:LioSpec} where we interpret the success of our proposal in terms of a decomposition of the Liovillian.

\section{Acknowledgements}
A.\,H.\,K. acknowledges financial support from the Villum Foundation and from the Danish Ministry of Higher Education and Science.
A.D.P acknowledges the financial support from
the University of Florence in the framework
of the University Strategic Project Program 2015
(project BRS00215).

%

\end{document}